\documentclass[twoside]{ilcws08}
\usepackage[latin1]{inputenc}
\usepackage[dvips]{graphicx,epsfig,color}
\usepackage{wrapfig,rotating}
\usepackage{amssymb,amsmath,array}
\usepackage{hyperref}

\begin{document}
\title{
Threshold corrections to the $\gamma t\bar{t}$ vertex at ${\cal O}(\alpha\alpha_s)$}

\author{Dirk Seidel
\thanks{This work is supported by the Alberta
Ingenuity foundation and NSERC. The slides of the talk can be obtained from~\cite{url}.}
\vspace{.3cm}\\
Department of Physics, University of Alberta\\
Edmonton, AB T6G 2J1, Canada
}

\maketitle

\begin{abstract}
In these proceedings a recent calculation of the last missing piece of the two-loop ${\cal
O}(\alpha \alpha_s)$ corrections to $\gamma t \bar{t}$ vertex at the $t \bar{t}$ threshold due
to the exchange of a $W$ boson and a gluon is summarised. The calculation constitutes a
building block of the top quark threshold production cross section at electron positron
colliders.
\end{abstract}

\section{Introduction}

The measurement of the total $t\bar t$ cross section constitutes a major part of the top
physics program at a future $e^+e^-$ International Linear Collider (ILC). Due to the large top
quark mass and the large width $\Gamma_t\gg\Lambda_{QCD}$, non perturbative effects are
strongly suppressed at energies around the threshold. Therefore it is possible to determine top
quark properties like the mass $m_t$ and the width $\Gamma_t$, but also the strong coupling
$\alpha_s$ with high precision. In Ref.~\cite{Martinez:2002st} it has been shown that an
uncertainty of below 100 MeV can be obtained for $m_t$ from a threshold scan of the cross
section.

The feasibility of such high-precision measurements requires a theory prediction of the total
cross section $\sigma(e^+ e^-\rightarrow t\bar{t})$ with high accuracy (preferably
$\delta\sigma/\sigma\leq 3\%$). The study of $t\bar{t}$ threshold production is performed in
the framework of non-relativistic QCD (NRQCD)~\cite{Bodwin:1994jh} which separates the hard and
soft scales involved in the process. The next-to-next-to-leading order (NNLO)
calculation~\cite{Hoang:2000yr} turned out to be as large as the NLO one. Only the (partial)
NNNLO~\cite{Beneke:2007pj,Beneke:2008ec} results show a good convergence of the perturbative
expansion and a strong reduction of the scale dependence from NNLO to NNNLO can be observed.
The remaining uncertainty turns out to be of the order of 10\%. A similar value is obtained
from the approach based on the resummation of logarithmically enhanced terms which have been
considered in Refs.~\cite{Hoang:2003xg,Pineda:2006ri}.

In order to reach a theory goal of $\delta\sigma/\sigma\leq 3\%$ it is necessary to include in
the prediction next to the one-loop electroweak corrections, which are known since quite some
time~\cite{Guth:1991ab} (see also~\cite{Hoang:2006pd}), also higher order effects. The
evaluation of ${\cal O}(\alpha \alpha_s)$ corrections has been started in
Ref.~\cite{Eiras:2006xm}, where the two-loop mixed electroweak and QCD corrections to the
matching coefficient of the vector current has been computed due to a Higgs or $Z$ boson
exchange in addition to a gluon. In these proceedings the calculation of ${\cal
O}(\alpha\alpha_s)$ for the two-loop vertex diagrams mediated by a $W$ boson and gluon
exchange is summarised~\cite{Kiyo:2008mh}. This result completes the vertex corrections of
order $\alpha\alpha_s$ --- a building block for the top quark production cross section.
Assuming the (numerically well justified) power counting $\alpha \sim \alpha_s^2$ one can see
that these corrections are formally of NNNLO.

In order to complete the matching corrections of order $\alpha\alpha_s$ also the two-loop box
diagrams contributing to $e^+ e^-\to t\bar{t}$ have to be considered. Actually, only the proper
combination of the box, vertex and self-energy contributions (the latter can, e.g., be found in
Refs.~\cite{Kniehl:1989yc,Djouadi:1993ss}) forms a gauge independent set.

\section{Threshold cross section}

Within the framework of NRQCD~\cite{Bodwin:1994jh} the total cross section for the top quark
production can be cast in the form
\begin{eqnarray}
  R\left(e_{L}^+ e_{R}^-  \rightarrow t\bar{t} X\right)
  &=&
   \frac{8\pi}{s}\,
   {\rm Im}
   \big[\, \left(h_{R,V}\right)^{\,2} H_{V}
           +\left(h_{R,A}\right)^{\,2} H_{A} \big]\,,
\label{eq:XSection}
\end{eqnarray}
where $s$ is the square of the centre-of-mass energy and $R(e_{L}^+ e_{R}^-  \rightarrow
t\bar{t} X)$ is the cross section normalised to $\sigma(e^+e^-\rightarrow
\mu^+\mu^-)=(4\pi\alpha^2)/(3s)$. For illustration we consider in Eq.~(\ref{eq:XSection})
left-landed positrons and right-handed electrons. For $e^+_R e^-_L$ in the initial state a
similar expression is obtained by replacing R by L in Eq.~(\ref{eq:XSection}). Note that in the
SM the initial states $e^+_R e^-_R$ and $e^+_Le^-_L$ are suppressed by a factor
$(m_e/M_W)^2\sim 10^{-10}$ and are thus negligible. We denote $h_{R,V}$ and $h_{R,A}$ by
helicity amplitudes which absorb the matching coefficients representing the effective coupling
of the effective operators. They take care of the hard part of the reaction. The first
subscript of $h$ refers to helicity of the electron, and the second one to the vector
($J_{V}^\mu=\bar{\psi}\gamma^\mu\psi$) or axial-vector coupling
($J_{A}^\mu=\bar{\psi}\gamma^\mu\gamma_5\psi$) of the gauge bosons to the top quark
current. The bound-state dynamics is contained in the so-called
hadronic part which is denoted by $H_V$ and $H_A$ in Eq.~(\ref{eq:XSection}).

In the cross section formula (\ref{eq:XSection}) ``${\rm Im}$'' refers to those cuts which
correspond to the $t\bar{t}X$ final state. This means that we have to select special cuts which
correspond to the final state we are interested in. This requires a dedicated study
incorporating the experimental setup. For the one-loop electroweak correction this treatment
was performed in \cite{Hoang:2004tg}. In~\cite{Kiyo:2008mh} we did not pursue this problem
further.

The evaluation of $H_V$ and $H_A$ requires to integrate out the low-energy modes of QCD, the
soft, potential and ultrasoft gluons contained in NRQCD \cite{Luke:1997ys,Beneke:1997zp}. For
the top quark system this can be done perturbatively. In a first step one integrates out the
soft and potential gluons which results in the effective field theory Potential
NRQCD~\cite{Pineda:1997bj,Brambilla:1999xf}. The 
\begin{wrapfigure}{r}{0.5\columnwidth}
\centerline{\includegraphics[width=0.45\columnwidth]{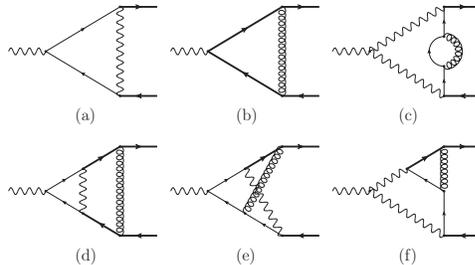}}
\caption{Sample Feynman diagrams contributing to the matching coefficient of the vector
current
at order $\alpha$ (a), $\alpha_s$ (b) and $\alpha\alpha_s$ (c)-(f). The thick (thin) straight
lines represent top (bottom) quarks, wavy lines stand for $W$ bosons and the curly ones for
gluons.}
\label{fig::sampleFDs} 
\end{wrapfigure}
corresponding Lagrangian is known to
NNNLO~\cite{Kniehl:2002br}.\footnote{The only missing constant in Ref.~\cite{Kniehl:2002br} is
related to the three-loop static potential where recently the fermion corrections became
available~\cite{Smirnov:2008pn}.} To integrate out nonrelativistic top and anti-top quark
fields the Rayleigh-Schr\"odinger perturbation theory can be applied as was initiated in
Ref.~\cite{Fadin:1987wz} and performed to NNNLO for $H_V$ and $H_A$ in
Refs.~\cite{Beneke:2008ec} and~\cite{Penin:1998mx}, respectively. Integrating out the ultrasoft
gluon was completed recently in Ref.~\cite{Beneke:2007pj}. For the details of these steps we
refer to the original papers and references cited therein (see also
Refs.~\cite{Kniehl:1999ud,Manohar:2000kr,Kniehl:2002yv,Hoang:2003ns,Penin:2005eu}).

In these proceedings we report on a recent publication, where corrections of ${\cal
O}(\alpha\alpha_s)$ to $h_{I,F}$ were computed. Some sample vertex diagrams which have to be
evaluated at threshold are shown in Fig.~\ref{fig::sampleFDs}. Let us mention that the
corresponding two-loop QCD corrections have been presented in
Refs.~\cite{Czarnecki:1997vz,Beneke:1997jm} and the three-loop corrections induced by a light
quark loop in Ref.~\cite{Marquard:2006qi}.

\section{Calculation of the vertex diagrams}

The two-loop diagrams which have to be evaluated are generated with {\tt
QGRAF}~\cite{Nogueira:1991ex} and further processed with {\tt q2e} and {\tt
exp}~\cite{Harlander:1997zb,Seidensticker:1999bb}. The reduction of the integrals is performed
with the program {\tt crusher}~\cite{PMDS} which implements the Laporta
algorithm~\cite{Laporta:1996mq,Laporta:2001dd}. We arrive at 29 master integrals (MI). Some MIs
factorise into one-loop integrals or contain only one dimensionful scale. Most of these
integrals are available in the literature and can, e.g., be found in
Refs.~\cite{Scharf:1993ds,Fleischer:1998dw,Seidel:2004jh,Kalmykov:2006pu,Eiras:2006xm,
Piclum:2007an}. 

The two-scale MIs are evaluated via an expansion in $z=M_W^2/m_t^2$. A promising method to
obtain the results is based on differential equations (see Ref.~\cite{Argeri:2007up} for a
recent review) which provide the expansion in an automatic way once the initial conditions are
specified. With the help of the ansatz
\begin{eqnarray}
  \text{MI} = \sum c_{ijk}\,\epsilon^i z^j (\ln z)^k,
\end{eqnarray}
the differential equations can be expanded in $\epsilon$ and $z$. As a result they reduce to
algebraic equations for the coefficients $c_{ijk}$, which can be solved trivially. In every
order in $\epsilon$ there is one constant $c_{ijk}$ which can not be determined with this
procedure. It is obtained from the initial condition at $z=0$. Unfortunately, we could not get
analytic results for five coefficients in the $\epsilon$-expansion of four MIs at $z=0$. We
calculated these coefficients using the Mellin-Barnes method (see, e.g.,
Ref~\cite{Smirnov:2004ym}) where we used the program packages {\tt AMBRE}~\cite{Gluza:2007rt}
and {\tt MB}~\cite{Czakon:2005rk}.

\begin{wrapfigure}{r}{0.35\columnwidth}
\centerline{\includegraphics[width=0.25\columnwidth]{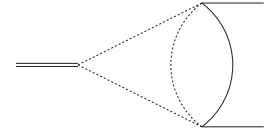}}
\caption{Solid(dotted) lines correspond to propagators with mass $m_t$(0). Single
external lines are on the mass shell with mass $m_t$ whereas double external line have mass
$2m_t$.}
\label{fig::MI} 
\end{wrapfigure}

The Mellin-Barnes representation for a given integral is not unique. In particular it might
happen that the convergence of the resulting numerical integration turns out to be good in one
case whereas a poor convergence is observed in other cases. The crucial quantity in this
respect is the asymptotic behaviour of the $\Gamma$ function for large imaginary part which is
given by
\begin{eqnarray}
  \Gamma(a\pm ib)\stackrel{\tiny b\to\infty}{\simeq}
  \sqrt{2\pi}e^{\pm i\frac{\pi}{4}(2a-1)}e^{\pm ib(\ln
    b-1)}e^{-\frac{b\pi}{2}}b^{a-\frac{1}{2}}
  \,,
  \label{eq::asymgamma}
\end{eqnarray}
where the first two exponential factors lead to oscillations. Let us discuss this in more
detail for the Mellin-Barnes representation of one of the MIs, which is depicted in
Fig.~\ref{fig::MI}:
\begin{eqnarray}
 \text{MI} &=& \left(\frac{e^{\gamma_E}
\mu^2}{m_t^2}\right)^{2\epsilon}
\int_{-i\infty}^{+i\infty} \frac{dz_1}{2\pi
i}\int_{-i\infty}^{+i\infty} \frac{dz_2}{2\pi i} \, 
 e^{i\pi(2 \epsilon+z_1)} 4^{-2 \epsilon-z_2}\nonumber\\
&&\hspace{-0em}\times  \frac{\Gamma
(1-\epsilon) \Gamma
   (-\epsilon-z_1+1) \Gamma (-z_1)\left[\Gamma (-2 
\epsilon-z_2+1)\right]^2 }{\Gamma (-3
   \epsilon-z_1+2)
   \Gamma (-2 \epsilon-z_1+2) \Gamma (-4 \epsilon-2
   z_2+2)}\nonumber\\
 &&\hspace{-0em}\times \Gamma
 (-4 \epsilon-z_1-z_2+2) \Gamma (z_1-z_2) \Gamma
 (\epsilon+z_2) \Gamma (2 \epsilon+z_2)
 \,,
 \label{eq::MB45}
\end{eqnarray}
where $\gamma_E=0.577216\dots$ and the contour of integration is chosen in such a way that the
poles of the $\Gamma$ functions with $+z_i$ are separated from the poles of the $\Gamma$
functions with $-z_i$. Using the package {\tt MB} we can expand the integrand in $\epsilon$.
For the finite contribution this leads to a sum of an analytic part, a one-dimensional
Mellin-Barnes integral and a two-dimensional one. The latter correspond to the integral in
Eq.~(\ref{eq::MB45}) for $\epsilon=0$. If we insert in this expression the asymptotic behaviour
for the $\Gamma$ functions as given in Eq.~(\ref{eq::asymgamma}) one observes that the
integrand of the two-dimensional integral falls off exponentially, except for
$\mbox{Im}(z_2)=0, \mbox{Im}(z_1)<0$.\footnote{ Note that for these values of $z_1$ the
exponential factor in the integrand of (\ref{eq::MB45}) increases exponentially.} On this line
the drop-off only shows a power-law behaviour which is dictated by the last factor of
Eq.~(\ref{eq::asymgamma}). In our particular case the drop-off turns out to be extremely slow
for the integration contour chosen by {\tt MB} which corresponds to $\mbox{Re}(z_1)=-1/4$ and
$\mbox{Re}(z_2)=-1/2$. Thus it is hard to get an accurate result by the numerical integration
since a highly oscillating functions has to be integrated. A closer look to the fall-off
behaviour in Eq.~(\ref{eq::asymgamma}) shows that it is possible to improve the drop-off for
Im$(z_1)\to-\infty$ by taking residues of the integrand in $z_2$ and thus shifting the
integration contour for $z_2$ more and more to positive values for Re$(z_2)$. In this way the
integrand becomes well-behaved and can be integrated numerically with sufficiently high
precision.

It has already been pointed out in~\cite{Czakon:2005rk} that for certain kinematical
configurations the Mellin-Barnes integrals exhibit poor convergence behaviour.  We  have shown
that at least for the threshold integrals needed in our calculation, it is possible to choose
the integration paths in a way to make the numerical integration possible. One may hope that
with the approach described here it will turn out to be possible to use Mellin-Barnes
integration for other troublesome integrals, too.

The other integrals which were solved using the Mellin-Barnes method show similar properties
than the MI above. In all cases it is possible to end up with integrals which could be
integrated numerically. The accuracy for the finite part of these integrals is sufficient to
obtain the final result with four significant digits.

Note that contrary to the default settings of {\tt MB} we do not use
{\tt Vegas} for 
the multidimensional numerical integrations. Instead we use {\tt Divonne}
which is available from the Cuba library~\cite{Hahn:2004fe}. 
For the integrals we have considered it leads to
more accurate results using less CPU time. 

We have performed an independent check of the initial conditions for
all the MIs employing the method of sector decomposition.
In particular we used the program {\tt FIESTA}~\cite{Smirnov:2008py}.

\section{Results}

The hard part of the cross section in~(\ref{eq:XSection}) at tree level is given by
\begin{eqnarray}
  h_{I,V/A}^{\rm tree} &=& 
  Q_e Q_t 
  + \frac{s \,\beta_I^{\,e} \,\beta_{V/A}^{\,t}}{s-M_Z^2}~~~
  {\rm with}~~ \beta_{V/A}^{\,t}=\frac{\beta_R^{\,t}\pm\beta_L^{\,t}}{2},
\label{eq:hamp}
\end{eqnarray}
where the $\beta_{I}^{\,f}$ is the coupling of a fermion ($f=e, t$) with helicity $I$ to the
$Z$ boson~\cite{Kiyo:2008mh}. We denote by $\Gamma_A^{\,t}$ the contribution of the sum of all
one-particle-irreducible diagrams to the $\gamma t\bar{t}$ vertex and parametrise the
radiative corrections in the form
\begin{eqnarray}
  \hat{\Gamma}_{A}^{\,t}=Q_t
  +{\hat{\Gamma}}_{A}^{\,t,\,(0,1)}
  +{\hat{\Gamma}}_{A}^{\,t,\,(1,0)}
  +\hat{\Gamma}_{A}^{\,t,\,(1,1)}
  \,,  
  \label{eq::Ghat}
\end{eqnarray}
where the hat denotes renormalised quantities and the subscripts $(i,j)$ indicate
corrections of ${\cal O}(\alpha^i\alpha_s^j)$. Substituting $\hat{\Gamma}_{A}^{\,t}$ for the
$Q_t$ in Eq.~(\ref{eq:hamp}) and retaining the relevant orders in the electroweak and strong
couplings leads to the corrections to the helicity amplitudes, $h_{I,V}^{(0,1)},
h_{I,V}^{(1,0)}$ and $h_{I,V}^{(1,1)}$. We further decompose $\hat{\Gamma}_{A}^{\,t}$ (and
similarly the quantities on the right-hand side of Eq.~(\ref{eq::Ghat})) according to the
contributions from the Higgs, $Z$ and $W$ boson exchanges:
\begin{eqnarray}
  \hat{\Gamma}_{A}^{\,t}= \hat{\Gamma}_{A,H}^{\,t} + \hat{\Gamma}_{A,Z}^{\,t} +
  \hat{\Gamma}_{A,W}^{\,t} 
  \,.
\end{eqnarray}
The results for the electroweak corrections read~\cite{Guth:1991ab,Eiras:2006xm,Kiyo:2008mh}
\begin{eqnarray}
  &&
  \hat{\Gamma}_{A, H}^{t,(1,0)} = 21.1 \times 10^{-3} ~ (10.6\times 10^{-3} )
  ~ ~ ~\mbox{for}~~ M_H=120 ~ (200)~{\rm  GeV}\,,
  \nonumber \\ &&
  \hat{\Gamma}_{A, W}^{t,(1,0)}=3.0 \times 10^{-3}\,,
  \nonumber \\ &&
  \hat{\Gamma}_{A, Z}^{t,(1,0)}=1.7 \times 10^{-3}\,,
  \nonumber \\
  &&
  \hat{\Gamma}_{A, H}^{t,(1,1)}
  =-17.6 \times 10^{-3} ~ (-6.6\times 10^{-3} ) 
  ~ ~ ~\mbox{for}~~ M_H=120 ~ (200)~{\rm  GeV}\,,
  \nonumber \\ &&
  \hat{\Gamma}_{A, W}^{t,(1,1)}=0.2\times 10^{-3}\,,
  \nonumber \\ &&
  \hat{\Gamma}_{A, Z}^{t,(1,1)}=-1.0\times 10^{-3}\,.
  \label{eq::numG}
\end{eqnarray}
One observes quite small corrections from the $W$ and $Z$ boson induced contributions. From
Eq.~(\ref{eq::numG}) one can read off that relatively big one-loop effects are obtained for
light Higgs boson masses. However, there is a strong cancellation between the one- and two-loop
terms resulting in corrections which have the same size as the sum of the one- and two-loop
contributions of the $W$ and $Z$ boson diagrams. In general moderate effects are observed
suggesting that in the electroweak sector perturbation theory works well, which is in contrast
to the pure QCD corrections. Let us in the end mention that $\hat{\Gamma}_{A, W}^{t,(1,1)}$
leads to a correction of $0.1\%$ to $R$.

\section{Acknowledgements}

Y. Kiyo and M. Steinhauser are gratefully acknowledged for the collaboration. Diagrams have
been drawn with Axodraw/Jaxodraw~\cite{Vermaseren:1994je,Binosi:2008ig}.


\begin{footnotesize}

\end{footnotesize}


\end{document}